\documentclass[11pt]{article}
\usepackage{amssymb}
\usepackage{amsmath}
\usepackage{graphicx}
\usepackage{soul}  % to use \st{} command to strikeoout text
\usepackage{listings} %  automatizing lining in verbatim

\usepackage{float}
\usepackage{fancyvrb}
\usepackage{xcolor}
\usepackage{subcaption}
\usepackage{hyperref}

\newtheorem{theorem}{Theorem}
\newtheorem{example}{Example}
\newtheorem{definition}{Definition}

\usepackage{pst-plot}
\usepackage{tikz}
\usetikzlibrary{arrows}
\usetikzlibrary{er,positioning}
\usetikzlibrary{patterns}
\usepackage[shortlabels]{enumitem}

\usepackage{cancel}

\newcommand{\Id }{\mbox{Id} \,}
\newcommand{\trace }{\mbox{tr} \,}
\newcommand{\homo }{\mbox{cong}}

\def\reals{\mathbb{R}}
\def\RS{${\cal RS}$ }

\newcommand{\comment}[1]{\mbox{}}

\def\qed{{\hfill{\vrule height5pt width3pt depth0pt}\medskip}}

\title{Central configurations in  the spatial  $n$-body problem for $n=5,6$ with equal masses }

\author{Ma{\l}gorzata Moczurad and Piotr Zgliczy\'nski\footnote{Partially supported by the NCN grant 2015/19/B/ST1/01454}\\
   \{malgorzata.moczurad, piotr.zgliczynski\}@ii.uj.edu.pl \\
Faculty of Mathematics and Computer Science, Jagiellonian University,\\
ul. prof. Stanis\l awa \L ojasiewicza 6,
30-348 Krak\'ow, Poland
}

\tikzstyle{io} = [fill=black,inner sep=2pt,circle]

\begin{document}

\maketitle

\begin{abstract}
We present a computer assisted proof of the full listing of central configurations for spatial $n$-body problem for $n=5$ and 6, with equal masses. For each central configuration  we give a full list of its euclidean  symmetries. For all masses sufficiently close to the equal masses
case we give an exact count of configurations in the planar case for $n=4,5,6,7$ and in the spatial case for $n=4,5,6$.
\end{abstract}

\tableofcontents

%===================================
\section{Introduction}

A central configuration\index{central configuration}, denoted as CC, is an initial configuration $(q_1, \ldots, q_n)$ in the Newtonian $n$-body problem, such that if the particles were all released with zero velocity, they would collapse toward the center of mass $c$ at the same time.
In the planar case, CCs are initial positions for periodic solutions which preserve the shape of the configuration. CCs also play an important role in the study of the topology of integral manifolds in the $n$-body problem (see~\cite{MSch} and the references given there).

%------------------------------------------------
\subsection{State of the art}

The investigation  of central configurations for equal masses is a subcase of more general problem of central configurations with arbitrary positive masses.
The general conjecture of finiteness of central configurations (relative equilibria) in the $n$-body problem is stated in \cite{Wintner} and appears as the sixth
problem of Smale's eighteen problems for the 21st century \cite{SmNext}. We refer the reader to our paper \cite{MZ} for the description of  the state of the art of the planar problem.

 For the spatial  $5$-body problem Moeckel in~\cite{M01}  established the generic finiteness of  Dziobek's CCs (CCs which are non-planar). A computer assisted work by Hampton and Jensen~\cite{HJ}  strengthens this result by giving an explicit list of conditions for exceptional values of masses.

The spatial 5-body problem with equal masses was studied by
Kotsireas and his coworkers (\cite{Kotsireas,FK,KL} and references given there). They have shown that possible CCs with some reflectional symmetry
can be divided into five classes. For  four of them they proved  (computer assisted proof) the existence of a CC and stated a hypothesis that no CC exists in the remaining one.   This hypothesis was later proven in \cite{ADL} through another  computer assisted proof. In our paper, we confirm that  there are exactly four non-equivalent non-planar CCs having a reflection symmetry for $n=5$ bodies with equal masses; moreover we prove that there
are no non-symmetric non-planar CCs for $n=5$.

Next relevant work on spatial CC for $n=5$ is~\cite{LS}, where a complete classification of the isolated CCs of the 5-body problem was given. The approach has a numerical component, hence it cannot be claimed  fully rigorous (on the other hand the existence of identified isolated CC, has been proven using the Krawczyk's operator, i.e.\ a tool from interval arithmetic we also use).
The proof also does not exclude the possibility that a higher dimensional
set of solutions exists.

The above  mentioned works study the polynomial equations derived from the equations for CC  using the (real or complex) algebraic geometry tools.
In contrast, we take a different approach: we use standard interval arithmetic tools, hence in principle we can treat also other potentials which cannot be reduced to polynomial equations.

%------------------------------------------------
\subsection{Main result}\label{sec:main-result}
In the paper, we show that there exists only a finite  number of different CCs (up to euclidean symmetries, scaling and permutations of bodies), for $n=5,6$ in the spatial $n$-body Newtonian problem with equal masses.
There are four non-planar CCs for $n = 5$ and nine for $n = 6$; they are listed in Section~\ref{sec:spatialCC}.
Moreover, for each of these CCs we give a list of all euclidean symmetries. In particular we show that each of these
 central configurations has some reflectional symmetry.

Hence any CC can be obtained from one of the above CCs by a suitable composition of translation, scaling, rotation, reflection and permutation of bodies.

\begin{theorem}
\label{thm:main}
In  the spatial $n$-body Newtonian problem with equal masses there exists exactly four for  $n=5$  and nine for $n=6$  types of non-planar CCs.
Each of them has a reflectional symmetry.
\end{theorem}

The method used in the present paper is a straightforward extension of our work \cite{MZ} on the planar case to the spatial one.  This is basically a brute force approach using standard interval arithmetic tools.

%------------------------------------------------
\subsection{Structure of the paper}
The paper is organized as follows. In Section~\ref{sec:cc-eq} and Section~\ref{sec:red-sys-equiv} we recall notations, definitions and fundamental equations for central configurations. In Section~\ref{sec:sym} we explain the method of finding and recognizing symmetries of CCs -- the idea is the same as in~\cite{MZ}, however algebraic details are different. We present a subject more formally, since symmetries in 3D are more complicated than in the planar case. In Section~\ref{sec:cap} we give some details of the computer assisted proof focusing on the reduction of the configuration space to be searched.
In Section~\ref{sec:spatialCC} we give a full listing of central configurations for $n=5$ and 6.
Section~\ref{sec:iso}  we count the number of non-equivalent CCs for the case of masses close to the equal mass case, where, informally speaking, we call two CCs equivalent if they have the same geometrical shape. In Section~\ref{sec:stable-planar} we report on our non-rigorous computations concerning  the instability of all planar CCs in the equal mass case.

%===================================
\section{Equations for central configurations}
\label{sec:cc-eq}

This section is almost identical to some parts of Section 2 in \cite{MZ}. It is included here just to make the paper reasonably self-contained.

By $|z|$ we denote the Euclidian norm of $z$, i.e.\ $|z|=\left(\sum_{i=1}^d z_i^2\right)^{1/2}$, where $z\in \mathbb{R}^d$.  By $(x|y)$ we denote the standard scalar product, i.e. $(x|y)=\sum_{i=1}^d x_iy_i$, where $x,y \in \mathbb{R}^d$. We often use $z^2:=(z|z)$.
Let $q_i \in \mathbb{R}^d$, $i=1,\dots,n$ and $d\geqslant 1$\index{$q_i$} (the physically interesting cases are $d=1,2,3$), where $q_i$ is a position of $i$-th body
with mass $m_i \in \mathbb{R}_+$\index{$m_i$}.
Let us set
\begin{equation}
  M=\sum_{i=1}^n m_i.\index{M} \label{eq:def-mass-sum}
\end{equation}
\emph{Central configurations}\index{central configurations} are solutions of the following system of equations (see~\cite{MSch}):
\begin{equation}
  \lambda (q_i-c) = \sum_{\substack{j=1\\
j\neq i}}^n \frac{m_j}{r_{ij}^3}(q_i - q_j)=:\frac{1}{m_i}f_i(q_1,\dots,q_n), \quad i=1,\dots,n \label{eq:cc-with-lambda}
\end{equation}
where $\lambda\in\reals$ is a constant, $c=\left(\sum_{i=1}^n m_i q_i\right)/M$ is center of mass,  $r_{ij}=r_{ji}=|q_i - q_j|$\index{$r_{ij}$} is the Euclidean distance between $i$-th and $j$-th bodies and
$(-f_i)$\index{$f_i$} is the force which acts on $i$-th body resulting from the gravitational pull of other bodies.
The system of equations (\ref{eq:cc-with-lambda}) has the same symmetries as the $n$-body problem. It is invariant with respect to group of
isometries of $\mathbb{R}^d$ and the scaling of variables.

The system~(\ref{eq:cc-with-lambda}) has $dn$ equations and $dn + 1$ unknowns: $q_i \in \mathbb{R}^d$ for $i = 1, \ldots, n$ and $\lambda \in \mathbb{R}_+$.
The system has a $\mathcal{O}(d)$  and scaling symmetry (with respect to $q_i$'s and $m_i$'s), where $\mathcal{O}(d)$ is an orthogonal group in dimension $d$.
If we demand that $c=0$ (which is obtained by a suitable translation) and $\lambda=1$ (which can be obtained by scaling $q_i$'s or $m_i$'s) we obtain the equations (compare \cite{MSch,Mlect2014,AK})
\begin{equation}
  q_i= \sum_{j,j\neq i} \frac{m_j}{r_{ij}^3}(q_i - q_j)=:\frac{1}{m_i}f_i(q_1,\dots,q_n), \quad i=1,\dots,n.
  \index{$q_i$} \label{eq:cc-kart}
\end{equation}
It is easy to see that if (\ref{eq:cc-kart}) is satisfied, then $c=0$ and  (\ref{eq:cc-with-lambda}) also holds for $\lambda=1$.
A point $q = (q_1,\dots,q_n) \in \left(\mathbb{R}^d\right)^n$ is called a \emph{configuration}\index{configuration}.  If $q$ satisfies (\ref{eq:cc-kart}), then it is called
a \emph{normalized central configuration}\index{normalized central configuration} (abbreviated as CC)\index{CC}.
For the future use we introduce the function $F:\Pi_{i=1}^n\mathbb{R}^{d } \to \Pi_{i=1}^n\mathbb{R}^{d }$ given by
\begin{equation}\label{eq:vector-field}
  F_i(q_1,\dots,q_n) =  q_i - \sum_{j,j\neq i} \frac{m_j}{r_{ij}^3}(q_i - q_j), \quad i=1,\dots,n.
  \index{$F_i$}
\end{equation}
Then the system (\ref{eq:cc-kart}) becomes
\begin{equation}\label{eq:cc-abstract}
  F(q_1,\dots,q_n)=0.
  \index{$F$}
\end{equation}
It is well know that for any $(q_1,q_2,q_3,\dots,q_n)\in (\reals^d)^n$ holds
\begin{eqnarray}
\sum_{i=1}^n f_i&=&0, \label{eq:n-mom-con} \\
\sum_{i=1}^n f_i \wedge q_i & = & 0, \label{eq:n-angular-mom-con}
\end{eqnarray}
where $v \wedge w$\index{$v \wedge w$} is the exterior product of vectors, the result being an element of exterior algebra. If $d=2,3$ it can be interpreted as the vector product of $v$ and $w$ in dimension $3$.  The identities (\ref{eq:n-mom-con}) and (\ref{eq:n-angular-mom-con})
are easy consequences of the third Newton's law (the action equals reaction) and the requirement that the mutual forces between bodies are in direction of the other body.

Consider system (\ref{eq:cc-kart}). After multiplication of  $i$-th equation by $m_i$ and addition of all equations using (\ref{eq:n-mom-con}) we obtain (or rather recover)
the center of mass equation
\begin{eqnarray}
 \left(\sum_{i=1}^n m_i\right) c=\sum_i m_i q_i = 0. \label{eq:cc-cofmass}
\end{eqnarray}
We can take the equations for $n$-th body and replace it with (\ref{eq:cc-cofmass}) to obtain an equivalent system.
\begin{eqnarray}
  q_i&=& \sum_{j,j\neq i} \frac{m_j}{r_{ij}^3}(q_i - q_j)=:\frac{1}{m_i}f_i(q_1,\dots,q_n), \quad i=1,\dots,n-1, \label{eq:cc-kart-1n-1} \\
   q_n&=&-\frac{1}{m_n}\sum_{i=1}^{n-1} m_i q_i. \label{eq:cc-kart-n-th}
   \index{$q_i$}
\end{eqnarray}

In Section~\ref{sec:red-sys-equiv} we use (\ref{eq:n-angular-mom-con}) to define a reduced system of equations for CCs which does
not have the degeneracies present in system (\ref{eq:cc-kart}).

%===================================
\section{The reduced system of equations for CC }
\label{sec:red-sys-equiv}

The goal of this section is to derive a set of equations (the reduced system of equations), which gives all CCs, but the system
will no longer have $\mathcal{SO}(3)$-symmetry.  This section is an extension to $d=3$ of the results of Section 5 in \cite{MZ}, where the planar case $d=2$ has been treated.

%------------------------------------------------
\subsection{Non-degenerate solutions of full and reduced systems of equations}

Following Moeckel \cite{Mlect2014} we state the following definition.
\begin{definition}
  \label{def:non-deg-cc} We say that a normalized central configuration $q=(q_1,\dots,q_n)$ is \emph{non-degenerate}
  if the rank of $D\!F(q)$ is equal to $dn-\dim \mathcal{SO}(d)$. Otherwise the configuration is called \emph{degenerate}.
\end{definition}
The idea of the above notion of degeneracy is to allow only for the  degeneracy related to the rotational symmetry of the problem, because
by setting $\lambda=1$ in (\ref{eq:cc-with-lambda}) and keeping the masses fixed we removed the scaling symmetry.

We  write the  system (\ref{eq:cc-kart-1n-1}--\ref{eq:cc-kart-n-th}) obtained from (\ref{eq:cc-abstract}) after removing the $n$-th body using the center of mass equation (condition (\ref{eq:cc-cofmass})) as
\begin{equation}\label{eq:cc-abstract-red}
  F_{\mathrm{red}}(q_1,\dots,q_{n-1})=0
  \index{$F_{\mathrm{red}}$}
\end{equation}
where $F_{\mathrm{red}}: \Pi_{i=1}^{n-1}\mathbb{R}^{d} \to \Pi_{i=1}^{n-1}\mathbb{R}^{d} $.
Then it is easy to see that $q=(q_1,\dots,q_{n-1},q_n)$ is a non-degenerate central configuration iff the rank
of $D\! F_{\mathrm{red}}(q_1,\dots,q_{n-1})$ is $d(n-1)-\dim \mathcal{SO}(d)$.

%------------------------------------------------
\subsection{The reduced system \RS}\label{sec:red-sys}

Let $d=3$.
The fact that the system of equations (\ref{eq:cc-kart}) is degenerate  make this system not amenable for the use of standard interval arithmetic methods (see for example the Krawczyk operator) to rigorously count all possible solutions.
We need to remove the $\mathcal{SO}(3)$-symmetry and then hope that all solutions will be non-degenerate. In this section we present such reduction.

Let us fix $k_1,k_2 \in \{1,\dots,n-1\}$, $k_1 \neq k_2$ and consider the following set of equations
\begin{eqnarray}
  q_i &=&\frac{1}{m_i}f_i(q_1,\dots,q_n(q_1,\dots,q_{n-1})), \quad i\in \{1,\dots,n-1\}, i \neq k_1,k_2 \label{eq:cc-red-i} \index{$q_i$}\\
  x_{k_1} &=& \frac{1}{m_{k_1}}f_{k_1,x}(q_1,\dots,q_n(q_1,\dots,q_{n-1})), \label{eq:cc-red-xk1}\\
   x_{k_2} &=& \frac{1}{m_{k_2}}f_{k_2,x}(q_1,\dots,q_n(q_1,\dots,q_{n-1})), \label{eq:cc-red-xk2}\\
    y_{k_2} &=& \frac{1}{m_{k_2}}f_{k_2,y}(q_1,\dots,q_n(q_1,\dots,q_{n-1})), \label{eq:cc-red-yk2}
\end{eqnarray}
where
\begin{eqnarray}
  q_n(q_1,\dots,q_{n-1})&=&-\frac{1}{m_n}\sum_{i=1}^{n-1} m_i q_i, \label{eq:cc-red-com}
\end{eqnarray}
where $f_i = (f_{i,x}, f_{i, y},f_{i,z})$.\index{$f_{k,x}$}
In the sequel, we use the abbreviation \RS to denote the reduced system (\ref{eq:cc-red-i}--\ref{eq:cc-red-yk2}).
Observe that \RS  coincides with the system (\ref{eq:cc-kart-1n-1}--\ref{eq:cc-kart-n-th}) with the
equations for $y_{k_1},z_{k_1},z_{k_2}$ dropped.

Observe also that \RS no longer has $O(3)$ as a symmetry group. But still it is symmetric with respect to the reflections against
the coordinate planes.

The next theorem addresses the question: whether from \RS we
obtain the solution of (\ref{eq:cc-kart})?

\begin{theorem}\label{thm:red-to-full}
\begin{description}
\item[Case 1]
   If $q = (q_1,\dots,q_n)$ is a solution of \RS and the following conditions are satisfied
    \begin{itemize}
    \item[(\!A\!1\!)] $x_{k_1} \neq x_n$,
    \item[(\!A\!2\!)] the vectors $(x_{k_1}-x_n,y_{k_1}-y_n)$ and $(x_{k_2}-x_n,y_{k_2}-y_n)$ are linearly independent,
    \end{itemize}
     then
    it is a normalized central configuration, i.e.\ it satisfies (\ref{eq:cc-kart}).

\item[Case 2]    If  $q$ is a solution of \RS such that $z_i=0$  for $i=1,\dots,n$ and condition (A1) is satisfied, then $q$  is a normalized central configuration, i.e. it satisfies (\ref{eq:cc-kart}).
\end{description}
\end{theorem}
\textbf{Proof:}
First we show Case 1.
For any configuration $q$ we set
\begin{equation}\label{eq:Ri}
  R_i(q_1,\dots,q_n)=m_iq_i - f_i(q_1,\dots,q_n), \quad i=1,\dots,n
\end{equation}
and for any $(q_1,\dots,q_{n-1}) \in (\mathbb{R}^3)^{n-1}$
we define
\begin{eqnarray}\label{eq:R-tylda}
  \tilde{R}_i(q_1,\dots,q_{n-1})=R_i(q_1,\dots,q_{n-1},q_n(q_1,\dots,q_{n-1})), \quad i=1,\dots,n.
\end{eqnarray}
We  use the notation  $\tilde{R}_i=(\tilde{R}_{i,x},\tilde{R}_{i,y},\tilde{R}_{i,z})$.
Observe that for any $(q_1,\dots,q_{n-1}) \in (\mathbb{R}^3)^{n-1}$  holds
\begin{eqnarray}
  \tilde{R}_n(q_1,\dots,q_{n-1})=-\sum_{i=1}^{n-1}\tilde{R}_n(q_1,\dots,q_{n-1}). \label{eq:c-mom-red}
\end{eqnarray}
Indeed, from (\ref{eq:n-mom-con}) and (\ref{eq:cc-red-com})--(\ref{eq:R-tylda}) we have
\begin{eqnarray*}
 \tilde{R}_n(q_1,\dots,q_{n-1}) & = & R_n(q_1,\dots,q_{n-1},q_n(q_1,\dots,q_{n-1})) \\
 & = &  m_n q_n(q_1,\dots,q_{n-1})-f_n(q_1,\dots,q_{n-1},q_n(q_1,\dots,q_{n-1}))  \\
 & = & -\sum_{i=1}^{n-1} m_i q_i + \sum_{i=1}^{n-1}f_i(q_1,\dots,q_{n-1},q_n(q_1,\dots,q_{n-1})) \\
 & = & - \sum_{i=1}^{n-1}\tilde{R}_n(q_1,\dots,q_{n-1}).
\end{eqnarray*}
Observe that from (\ref{eq:n-angular-mom-con}) it follows that for any configuration $(q_1,\dots,q_n)$ holds
\begin{equation}
  \sum_{i=1}^n q_i \wedge R_i(q_1,\dots,q_n)=0.
\end{equation}
In particular for
 $q_n=q_n(q_1,\dots,q_{n-1})$ we obtain from (\ref{eq:c-mom-red})
\begin{eqnarray}
  0 & = & \sum_{i=1}^{n-1}q_i \wedge \tilde{R}_i \left(q_1,\dots,q_{n-1}\right) + q_n(q_1,\dots,q_{n-1}) \wedge \tilde{R}_n \left(q_1,\dots,q_{n-1}\right) \nonumber\\
  & = &
  \sum_{i=1}^{n-1}(q_i-q_n) \wedge \tilde{R}_i(q_1,\dots,q_{n-1}). \label{eq:c-angmom-red}
\end{eqnarray}

Form now on we assume that $q=(q_1,\dots,q_{n-1})$ is a solution of \RS and $q_n=q_n(q_1,\dots,q_{n-1})$.
Without any loss of the generality we can assume that $k_1=n-1$ and $k_2=n-2$.
We need to show that $\tilde{R}_{n-1,y}(q)=\tilde{R}_{n-1,z}(q)=\tilde{R}_{n-2,z}(q)=0$.
From (\ref{eq:c-angmom-red}) we have
\begin{eqnarray*}
0 & = &\sum_{i=1}^{n-1}(q_i-q_n) \wedge \tilde{R}_i(q)\\
  & = & (q_{n-2}-q_n)\wedge \tilde{R}_{n-2}(q) + (q_{n-1}-q_n)\wedge \tilde{R}_{n-1}(q)\\
  & = & (q_{n-2}-q_n)\wedge (0,0,\tilde{R}_{n-2,z}(q)) + (q_{n-1}-q_n)\wedge (0,\tilde{R}_{n-1,y}(q),\tilde{R}_{n-1,z}(q))\\
 & = &
  \begin{bmatrix}
  (y_{n-2}-y_n)\tilde{R}_{n-2,z}(q) + (y_{n-1}-y_n)\tilde{R}_{n-1,z}(q) - (z_{n-1}-z_n)\tilde{R}_{n-1,y}(q)  \\
  -(x_{n-2}-x_n)\tilde{R}_{n-2,z}(q) - (x_{n-1}-x_n)\tilde{R}_{n-1,z}(q) \\
  (x_{n-1}-x_n)\tilde{R}_{n-1,y}(q)
\end{bmatrix}
\\
& = &
\begin{bmatrix}
  (y_{n-2}-y_n) &  - (z_{n-1}-z_n) &   (y_{n-1}-y_n)  \\
  -(x_{n-2}-x_n) & 0 & -(x_{n-1}-x_n)  \\
  0 & (x_{n-1}-x_n) & 0
\end{bmatrix}
\cdot
\begin{bmatrix}
   \tilde{R}_{n-2,z}(q) \\
     \tilde{R}_{n-1,y}(q) \\
       \tilde{R}_{n-1,z}(q)
\end{bmatrix}.
\end{eqnarray*}
This is a homogenous linear system.  If the determinant of its matrix is non-zero, then it has only the zero solution. It is easy to see that this implied
by the following two conditions
\begin{eqnarray*}
  x_{n-1}-x_n \neq 0 \\
  \det \left[\begin{array}{cc}
  (y_{n-2}-y_n), &   (y_{n-1}-y_n)  \\
  -(x_{n-2}-x_n),  & -(x_{n-1}-x_n)  \\
\end{array}
\right] \neq 0
\end{eqnarray*}
The second condition means that vectors $(x_{n-2}-x_n,y_{n-2}-y_n)$ and $(x_{n-1}-x_n,y_{n-1}-y_n)$ are linearly independent.

Now we treat Case 2, where $z_i=0$, for $i=1,\dots,n$.  Then we obtain  $\tilde{R}_{n-1,z}(q)=\tilde{R}_{n-2,z}(q)=0$  and
we just need to show that $\tilde{R}_{n-1,y}(q)=0$.  After substitution of this information in the above formulas we obtain
 \begin{eqnarray*}
0 & = &\sum_{i=1}^{n-1}(q_i-q_n) \wedge \tilde{R}_i(q)
 = \begin{bmatrix}
   0 \\
   0 \\
   (x_{n-1}-x_n) \tilde{R}_{n-1,y}(q)
\end{bmatrix}
\end{eqnarray*}
and our assertion follows immediately.
\qed

Observe that condition (A2) is never satisfied for collinear solutions and also might not be satisfied for some planar solutions containing
three collinear bodies - such solutions exist  for $n=5$ and more, see \cite[Sec. A.2]{MZ}. This is why we included the second assertion in Theorem~\ref{thm:red-to-full}.

Other issue is how to know that a particular solution of the reduced system  (\ref{eq:cc-red-i}--\ref{eq:cc-red-yk2}) is contained in the plane $\{z=0\}$, while our information about the solution is that it is a unique solution in some interval set (box), which is not contained in this plane. This issue is addressed below.

%------------------------------------------------
\subsection{Collinearity and coplanarity tests }
\label{subsec:co-tests}

To define reduced system we select $k_1=n-1$ and $k_2=1$, which implies  that $q_{n-1}=(x_{n-1},0,0)$ and $q_1=(x_1,y_1,0)$.
Let us denote by $R_y(x,y,z)=(x,-y,z)$ and  $R_z(x,y,z)=(x,y,-z)$ the reflections with respect to $OXZ$ and $OXY$ planes, respectively.
Let $q$ be a configuration. For any map $R:\mathbb{R}^d\to \mathbb{R}^d$ we set
\begin{equation*}
  R(q)=(Rq_1,\dots,R q_n).
\end{equation*}
The proposed tests are based on the effective procedure to check the local uniqueness for the reduced system which, in our case, is the application of the Krawczyk operator \cite{K} (see also \cite[Sec. 6.3]{MZ}).

%------------------------------------------------
\subsubsection{Coplanarity test}
Observe that if $q$ is a coplanar solution of the reduced system  and $q_1$, $0$ and $q_{n-1}$ are not collinear, then $q$ must be contained in the plane $\{z=0\}$.

\begin{theorem}
\label{thm:co-planar -test}
Assume that $C$ is a box containing a unique solution $q$ of \RS and
 that the set  $C \cup R_z C$ contains a unique solution of that system. Then $q=R_z q$, i.e. $q$ is contained in the plane $\{z=0\}$.
\end{theorem}
\textbf{Proof:}
Observe that \RS is symmetric with respect to $R_z$, hence if $q$ is a unique solution of \RS in $C$, then $R_z q$ is a unique solution of \RS in $R_z C$.
From the uniqueness in   $C \cup R_z C$ it follows that $q=R_z q$.
\qed

%------------------------------------------------
\subsubsection{Collinearity test}
Observe that if $q$ is a collinear solution of \RS\!\!, then it is contained in the OX-axis. Indeed, $q_{n-1}$ and $c=0$ (the center of mass) belongs to the line containing CC.

\begin{theorem}
\label{thm:co-lin-test}
Assume that $C$ is a box containing a unique solution of \RS
and that each of the sets $C \cup R_y C$ and $C \cup R_z C$ contains a unique solution of \RS\!\!, then the unique solution in $C$ is collinear.
\end{theorem}
\textbf{Proof:}
From the coplanarity tests it follows that $y_i=z_i=0$, for all $i=1,\dots,n$. Hence the solution is contained in the $OX$-axis.
\qed

%===================================
\section{Symmetries}
\label{sec:sym}

The goal of this section is to describe a method which allow to find all orthogonal symmetries of spatial configurations.
Conceptually this is the same as in \cite{MZ}. However the task of finding symmetries in 3D is a bit more involved, thus we are more formal this time and we devote a full section to it. In this section we index bodies from $0$ to $n-1$ to be in the agreement with the program.
Let us stress that in Section~\ref{subsec:co-tests} we describe an effective test, which tell us whether a solution of \RS
is coplanar (which means that it is contained in the plane $\{z=0\}$).
As in Section~\ref{subsec:co-tests} (modulo indexing bodies starting from $0$), we assume that \RS is defined by $k_1=n-2$ and $k_2=0$.

\begin{definition}
Let $\sigma:\{0,1,\dots,n-1\} \to \{0,1,\dots,n-1\}$ be a permutation (i.e.\ $\sigma \in S_n$) and  $R \in \mathcal{O}(3)$. Then $(R,\sigma)$ is an (orthogonal) symmetry of a configuration
$(q_0,\dots,q_{n-1})$ iff
\begin{equation}
   q_{\sigma(i)}=R q_i, \quad i=0,1,\dots,n-1.
\end{equation}
\end{definition}

\begin{definition}
For $\sigma \in S_n$ and $R \in \mathcal{O}(3)$ we define a map
\begin{equation*}
  (R,\sigma): (\mathbb{R}^3)^n \to  (\mathbb{R}^3)^n ,
\end{equation*}
by
\begin{equation}
  ((R,\sigma)q)_{\sigma(i)}=Rq_i, \quad i=0,1,\dots,n-1.
\end{equation}
\end{definition}
Obviously, $q$ has symmetry $(R,\sigma)$ iff $(R,\sigma)q=q$.

In \RS the special role is given to  the $O\!X$-axis (which contains $q_{n-2}$) and the plane $\{z=0\}$ (containing $q_0$ and $q_{n-2}$). This should be taken into account when looking for orthogonal matrix $R$ which is a symmetry of a given CC. It should act so that
from one solution of \RS we should obtain another solution of \RS.

Let us stress that in the paper a normalized central configuration $q$ is the unique solution of \RS in an interval set $Z$ and is represented by this interval set.
 The basic idea is first to find a good candidate for $R$ and then look for possible $\sigma$. Once we have such candidate $(R,\sigma)$, we take $Z'=\mbox{intervalHull}(Z,(R,\sigma)(Z))$ and show, using the Krawczyk method, the uniqueness of CC in $Z'$. From this it follows that $q$ and its symmetric image $(R,\sigma)q$, both being the solutions of the reduced system in $Z'$, coincide.

Below we describe a procedure, which allows to find all symmetries of a given CC.  We need to find both $R \in \mathcal{O}(3)$ and
$\sigma \in S_n$.

%------------------------------------------------
\subsection{Finding candidates for $(R, \sigma)$}

CC is given as interval sets $q_i \subset \mathbb{R}^3$, $i=0,1,\dots,n-1$.
We assume that
\begin{equation}
x_{n-2}>0, \quad y_0 >0.
\end{equation}
Moreover, we assume that $y_0 \geqslant |y_i|$, for $i=0,1,\dots,n-1$.
Observe that this condition implies that CC is not collinear.
In (\ref{sec:init})--(\ref{sec:constsigma}) below, we describe each step of construction of $(R,\sigma)$.

%------------------------------------------------
\subsubsection{Initialization of $\sigma$}\label{sec:init}
At the beginning, $\sigma(i)$ is undefined for all $i$.

%------------------------------------------------
\subsubsection{Finding candidates for symmetric images of $q_{n-2}$ and $q_0$}\label{sec:find-cand}

Since we need to check all possibilities we repeat the below procedure for each $i\in \{0, \ldots, n-1\}$:
\begin{itemize}
\item if $|q_i|\cap |q_{n-2}|\neq \emptyset$, then set $\sigma(n-2) = i$;
\item if $0 \in |q_i|$, then we would exit with failure; however this never happens in our program;
\item
let us define $e_x = q_i/|q_i|$ and $t_j = |q_j - (q_j|e_x)e_x|$. We look for $j\neq i$ such that $t_j \cap y_0 \neq \emptyset$.  If there is no such $j$, we abandon the construction and continue for the next $i$. Otherwise
we set $\sigma(0)=j$.
\end{itemize}
In this way we identify all possible images of $(n-2)$-th and $0$-th bodies by orthogonal symmetries.
Observe that if $\sigma(n-2)=n-2$ and $\sigma(0)=0$, then $R$  is an identity on  the plane $z=0$. i.e.\ $R$ is either an identity in $\reals^3$ or the reflection with respect to $z=0$ plane.

%------------------------------------------------
\subsubsection{Constructing $R_+$ and $R_-$ --- the candidates for the symmetries}\label{sec:constR}

At this moment, we have a candidate for the image of the plane $\{z=0\}$. This is a plane containing $0$, $q_{\sigma(n-2)}$ and $q_{\sigma(0)}$. If $0 \in |\tilde{e}_y|$, we abandon the construction and return failure (this never happens in the program, because we know that the solution is not collinear and  bounds on $q_i$ are tight).
Let us define $\tilde{e}_y=q_{\sigma(0)} - (q_{\sigma(0)}|e_x)e_x$ and $e_y=\tilde{e}_y/ |\tilde{e}_y|$.
We define $e_z=e_x \times e_y$. Now we have two possibilities for $R$, denoted by $R_\pm$. We define them on the standard basis $e_1,e_2,e_3$ as follows
\begin{eqnarray*}
  R_\pm(e_1)=e_x, \quad R_\pm(e_2)=e_y, \quad R_\pm(e_3)=\pm e_z.
\end{eqnarray*}
Observe that $\det R_+=1$ and $\det R_-=-1$.

%------------------------------------------------
\subsubsection{Construction of $\sigma$}\label{sec:constsigma}
Let us fix $R=R_+$ or $R=R_-$. We extend the definition of $\sigma$ by demanding that for each $i$ there exists unique $j=\sigma(i)$, such that
 \begin{equation*}
   q_{j} \cap R q_i \neq \emptyset.
\end{equation*}

%------------------------------------------------
\subsection{Geometric description of $(R,\sigma)$}
Once we know a symmetry  $(R,\sigma)$ of CC $q$, we want to recognize its  geometric features, for example the angle of rotation etc.

We have two possibilities for $R$: it is either a rotation around some axis or is an  improper rotation, which is composition of  rotation (we allow also for identity) with reflections and are characterized by orthogonal matrices with determinant $-1$.  We should stress that even though $R$ is given as an interval matrix, we can give exact value of the rotation angle, since
we have discrete
set of points $q_i$ which are permuted by $\sigma$.

%------------------------------------------------
\subsubsection{Rotations}
The eigenvalues of $R$ are  $\{1, e^{\pm i \varphi}\}$. The eigenvector corresponding to $1$ is the axis of rotation and in the perpendicular plane we have a rotation by the angle $\varphi$.

In order to determine $\varphi$ we decompose $\sigma$ into cycles. The cycles of length $1$ consist of
points on the rotation axis. All other cycles should be of the same length $k$ and the rotation angle is $\varphi=2\pi/k$. Observe that there must be $k>0$
because we assumed that configuration is  non collinear.

%------------------------------------------------
\subsubsection{Improper rotations}
 The eigenvalues of $R$ are $\{-1, e^{\pm i \varphi}\}$.
 We have three possibilities:
 \begin{enumerate}
 \item the eigenvalue $-1$ has multiplicity three. In such situation $R=-\Id$ and the decomposition of $\sigma$ into cycles
   has the following properties: {\em there exists at most one cycle of length $1$ (this must be $q_i=0$) and all  other cycles are of length two}.

 \item the eigenvalue $-1$ has multiplicity one and $1$ has the multiplicity two.  This is a reflection with respect to some plane. Thus the decomposition of $\sigma$ into cycles  has the following properties:
   \begin{itemize}
     \item {\em there might be several cycles of length $1$}, these are points on the reflection plane,
     \item {\em all other  cycles are of length $2$}, if the configuration is non coplanar then at least one such cycle should appear,
   \end{itemize}
 \item all eigenvalues have multiplicity one. The eigenvector corresponding to $-1$ is the axis of rotation and in the perpendicular plane we have a rotation by the angle $\varphi$. In this situation the decomposition of $\sigma$ into cycles    has the following properties:
     \begin{itemize}
     \item {\em there can be at most one cycle of length $1$}, this must a point at the origin.
    \item {\em there can be cycles of length $2$, located on the "rotation" axis},
    \item {\em all other cycles should be of the length $k$ or $2k$, where $\varphi=2\pi/k$}; observe that $k$ must be greater than $2$ ($k=2$ is taken care of in the case of  $-1$ having multiplicity three).
     \end{itemize}
\end{enumerate}
From the above consideration it is clear that looking on $\sigma$ alone we might not be able to distinguish the cases 1 and 2, but this can be done easily by additionally estimating the eigenvalues of $R$.  In the third case  we may need to compute eigenvalues to decide on the value of $k$.

To determine the angle $\varphi$ we use the fact that if $A$ is a linear operator represented by a square matrix and $\lambda_1, \ldots, \lambda_n$ are the eigenvalues of $A$, then
\begin{equation*}
\trace(A) = \sum_{i=1}^n \lambda_i =  \sum_{i=1}^n a_{ii}.
\end{equation*}
Since in the orientation reversing case the eigenvalues of $R$ are $\{-1, e^{\pm i \varphi}\}$, we obtain
\begin{eqnarray*}
\trace(R) & = & -1 + e^{i\varphi} + e^{-i\varphi}\\
  & = & -1 + 2\cos\varphi ,
\end{eqnarray*}
hence
\begin{equation}
  \cos \varphi= \frac{\trace (R) + 1}{2}.  \label{eq:TrR-cos}
\end{equation}
Moreover, if $R = -\Id$, then $\trace(R) = -3$ and if $R$ is a reflection with respect to some plane, then  $\trace(R)=1$.
Based on the above observations we recognize $R$ as follows:
\begin{itemize}
\item if all cycles in $\sigma$ are of length at most $2$, then
  \begin{itemize}
    \item if $\trace(R) < 1$, then $R=-\Id$
    \item if $\trace(R) > -3$, then $R$ is a refection with respect to the plane perpendicular to the vector $q_i-q_{\sigma(i)}$, where $i$ is such that $\sigma(i)\neq i$, i.e. we take any cycle of length two
    \item if neither of the above holds, then we cannot decide between $R=-\Id$ and $R$ being the reflection (this never happens in our computations)
  \end{itemize}
\item if $\sigma$ contains a cycle of length $k>2$, then we have either $\varphi=\frac{2\pi}{k}$ or $\varphi=\frac{2\pi}{k/2}=\frac{4\pi}{k}$.
  The correct value is obtained by testing the formula (\ref{eq:TrR-cos}), i.e.
  \begin{itemize}
  \item if only one of the following conditions holds
    \begin{eqnarray*}
    \cos\frac{2\pi}{k} \in \frac{\trace (R) + 1}{2} &  \quad\mbox{or}\quad &
   \cos\frac{4\pi}{k} \in \frac{\trace (R) + 1}{2},
   \end{eqnarray*}
   then we know the value of $\varphi$,
     \item otherwise we cannot decide between these two possibilities (this never happens in our computations).
  \end{itemize}
\end{itemize}

%===================================
\section{On the computer assisted proof}
\label{sec:cap}

We normalized masses so that $M=\sum_i m_i=1$.
As in the previous  section we index bodies from $0$ to $n-1$ to be in the agreement with the program.
In the sequel, we study the following \RS with $k_1=n-2$ and $k_2=0$ (this is the same choice as in the previous section)
\begin{eqnarray}
  q_i &=&\frac{1}{m_i}f_i(q_0,\dots,q_{n-2},q_{n-1}(q_0,\dots,q_{n-2})), \quad i\in \{1,\dots,n-3\},  \label{eq:cc-red-final-i} \index{$q_i$}\\
  x_{n-2} &=& \frac{1}{m_{n-2}}f_{n-2,x}(q_0,\dots,q_{n-2},q_{n-1}(q_0,\dots,q_{n-2})), \label{eq:cc-red-final-xk1}\\
   x_{0} &=& \frac{1}{m_{0}}f_{0,x}(q_0,\dots,q_{n-2},q_{n-1}(q_0,\dots,q_{n-2})), \label{eq:cc-red-final-xk2}\\
    y_{0} &=& \frac{1}{m_{0}}f_{0,y}(q_0,\dots,q_{n-2},q_{n-1}(q_0,\dots,q_{n-2})), \label{eq:cc-red-final-yk2}
\end{eqnarray}
where
\begin{eqnarray}
 y_{n-2}&=&0, \label{eq:cc-red-final-yn-2}  \\
  z_{n-2}&=&0, \label{eq:cc-red-final-zn-2}  \\
   z_{0}&=&0, \label{eq:cc-red-final-z0}  \\
  q_{n-1}(q_0,\dots,q_{n-2})&=&-\frac{1}{m_{n-1}}\sum_{i=0}^{n-2} m_i q_i, \label{eq:cc-red-final-com}
\end{eqnarray}

%------------------------------------------------
\subsection{Equal mass case, the reduction of the configuration space for CCs}
\label{subsec:cap-equal-masses}

After a suitable permutation of bodies and an orthogonal transformation it is easy to see that each CC has its equivalent
in the set of the configurations satisfying the following conditions
\begin{itemize}
\item $q_{n-2}=(x_{n-2},0,0)$ is the furthermost body from the origin, $x_{n-2}>0$,
\item $q_0=(x_0,y_0,0)$ is the point furthest from the line $O\!X$ (which is determined by $q_{n-2}$), $y_0 \geqslant 0$
\item $q_1=(x_1,y_1,z_1)$ is the point furthest from the plane $O\!X\!Y$ (which is determined by $q_{n-2}$ and $q_0$), $z_1 \geqslant 0$
\item all other bodies have their $x$ coordinates in the order of increasing/decreasing indices.
\end{itemize}
This, combined with  Theorem~9 and Lemma~10  in \cite{MZ}, shows that it is enough  to consider the following set in which we look for the central configurations
 \begin{eqnarray}
  0.5 \leqslant x_{n-2} & \leqslant & (n-1), \label{eq:con0} \\
  x_{n-2} &\geqslant& |q_i| , \quad i=0,\dots,n-1, \label{eq:xn>qi} \\
   x_{n-2} &\geqslant& |x_i|, \quad i=0,\dots,n-1, \label{eq:con1}\\
   y_{n-2} & = & z_{n-2} =  0\\
  y_0 &\geqslant& |y_i| , \quad i=0,\dots,n-1 \label{eq:con2}\\
 z_0 & = & 0\\
    z_1 &\geqslant& |z_i| , \quad i=0,\dots,n-1 \label{eq:con3} \\
  x_2 & \geqslant & x_3 \geqslant \dots \geqslant x_{n-3} \geqslant x_{n-1}.\label{eq:con4}
 \end{eqnarray}
We call this order {\em decreasing} due to the requirement~(\ref{eq:con4}).

%------------------------------------------------
 \subsection{Outline of the approach}
 In the algorithm we look for all zeros of the reduced system (\ref{eq:cc-red-final-i}--\ref{eq:cc-red-final-yk2}), which under assumptions (\!A\!1\!) and (\!A\!2\!)   of   Theorem~\ref{thm:red-to-full} is equivalent to (\ref{eq:cc-kart}) for the non coplanar solutions, while (\!A\!1\!) is sufficient for the coplanar ones.  These assumptions translate to the following conditions
 \begin{eqnarray}
   x_{n-2} &\neq& x_{n-1},  \label{eq:A1-spec} \\
    \det \begin{bmatrix}
     (x_0-x_{n-1})  & (x_{n-2}-x_{n-1})  \\
  (y_0-y_{n-1}) &   (y_{n-2}-y_{n-1})
\end{bmatrix}
&\neq& 0. \label{eq:A2-spec}
 \end{eqnarray}
Observe that condition (\ref{eq:xn>qi}) implies  (\ref{eq:A1-spec}).

In the program, we verify condition (\ref{eq:A2-spec}) computing the determinant for the non-planar solutions of the reduced system.

During the program, proving the existence of a locally unique solution in a box  is just as important as proving that there is no solution there.
Just as in \cite{MZ} for proving the existence  we use the Krawczyk operator applied to the system (\ref{eq:cc-red-final-i}--\ref{eq:cc-red-final-yk2}). To rule out the  existence of a solution we use the exclusion tests discussed in Section
 4 in \cite{MZ} and also the Krawczyk operator.

As in \cite{MZ} the collinearity, coplanarity and the symmetries of CCs are established by proving the uniqueness in a suitable symmetric box (see~Sections~\ref{subsec:co-tests} and \ref{sec:sym} for details).

%------------------------------------------------
\subsection{The algorithm}
\label{sec:alg}
The algorithm runs in the reduced configuration space which is a subset of $\reals^{3(n-1)-3}$, i.e.\ a configuration is represented by a point $(x_0, y_0,x_1,y_1,z_1,\ldots, x_{n-3}, y_{n-3},z_{n-3}, x_{n-2})$. Physically, we interpret such a configuration as the positions of  $n-1$ bodies with
$q_0=(x_0,y_0,0)$, $q_i=(x_i, y_i,z_i)$ for $i = 1, \ldots, n-3$ and $q_{n-2}=(x_{n-2},0, 0)$. From (\ref{eq:cc-red-final-com}) we obtain $q_{n-1}$ --- the position of the last body.

The algorithm is the same as for $d=2$, which was discussed in \cite{MZ}. The data types are essentially the same, with obvious modifications
taking into account the dimension of the space.

%------------------------------------------------
\subsection{Technical data}
\label{subsec:tech-data}
The main computations  were
carried out in parallel using the template function {\tt std::async} (from the standard C++ library) which runs the function  asynchronously (potentially in a separate thread which may be part of a thread pool) on Dell R930 4x Intel Xeon E7-8867 v3 (2,5GHz, 45MB), 1024 GB RAM. The compiler  is gcc version 4.9.2 (Debian 4.9.2-10+deb8u2).
Times obtained for different number of bodies are presented in Table~\ref{tab:cc}.

\begin{table}[h]
\begin{center}
\begin{tabular}{r|r|r| r }
no bodies & no CCs & total no of & elapsed time \\
 & & CPU-seconds & h:m:s.d \\
\hline
4 & 5 &  13.68 & 0:00:02.89 \\  % ASYNCH = 128
5 & 9 &   12502.35 & 0:10:56.00 \\ % ASYNCH = 256
6 & 18 & 103619048.67 & 534:28:05.76 \\
\end{tabular}
\caption {Comparison of execution times for different number of bodies. Third column  contains total of CPU times used by all threads during the run of the program; fourth  --- real or clock times.}\label{tab:cc}
\end{center}
\end{table}

%------------------------------------------------
\subsection{Source and references files}
Locations of all files related to the numerical part of the proof are specified in Table~\ref{tab:sourceFiles}; these include source code of the program with installation instructions,  two kinds of output files (text and binary) and Mathematica notebookes with CCs presentations.

Note that we use interval arithmetic, therefore positions of bodies, in both: text and binary output files, are given  as truncated intervals containing the true value. Positions of bodies used in Mathematica notebooks cannot be treated as reference; these notebooks are attached only for visualisation.

 In the text report files all non-planar CCs are given together with the symmetries description;
planar ones --- are indicated by numbering {\tt collinear solution no...} or  {\tt planar solution no ...}.

In binary files, we store just the positions of bodies. These are numbers written with the accuracy obtained by the program; in our case it is IEEE 754 double. The data  can be read independently --- see the function {\tt readBodiesFromBinaryFile} in the source code.
\begin{table}[h]
\begin{center}
\begin{tabular}{l|l}
\hline
\multicolumn{2}{c}{\bf Program and instructions}\\
\hline
source code & \href{http://www.ii.uj.edu.pl/~zgliczyn/papers/cc-3d/nbodies.zip}{\tt nbodies.zip}\\
\hline
installation instructions & \href{http://www.ii.uj.edu.pl/~zgliczyn/papers/cc-3d/Readme.html}{\tt Readme.html}\\
\hline
\hline
\multicolumn{2}{c}{\bf Output binary files}\\
\hline
five bodies & \href{http://www.ii.uj.edu.pl/~zgliczyn/papers/cc-3d/zeros-3D-5-1.dat}{\tt zeros-3D-5-1.dat}\\
 & \href{http://www.ii.uj.edu.pl/~zgliczyn/papers/CC-3D/undecided-3D-5-1.dat}{\tt undecided-3D-5-1.dat}\\
 \hline
six bodies & \href{http://www.ii.uj.edu.pl/~zgliczyn/papers/cc-3d/zeros-3D-6-1.dat}{\tt zeros-3D-6-1.dat}\\
 & \href{http://www.ii.uj.edu.pl/~zgliczyn/papers/cc-3d/undecided-3D-5-1.dat}{\tt undecided-3D-5-1.dat}\\
\hline
\hline
\multicolumn{2}{c}{\bf Text report files}\\
\hline
five bodies & \href{http://www.ii.uj.edu.pl/~zgliczyn/papers/cc-3d/cc-5b-3d.txt}{\tt cc-5b-3d.txt}\\
six bodies & \href{http://www.ii.uj.edu.pl/~zgliczyn/papers/cc-3d/cc-6b-3d.txt}{\tt cc-6b-3d.txt}\\
\hline
\hline
\multicolumn{2}{c}{\bf Mathematica notebooks}\\
\hline
five bodies & \href{http://www.ii.uj.edu.pl/~zgliczyn/papers/cc-3d/cc-5b-3d.nb}{\tt cc-5b-3d.nb}\\
six bodies & \href{http://www.ii.uj.edu.pl/~zgliczyn/papers/cc-3d/cc-6b-3d.nb}{\tt cc-6b-3d.nb}\\
\hline
\end{tabular}
\captionsetup{justification=centering,margin=2cm}
\caption{All links above refer to the directory: {\tt http://www.ii.uj.edu.pl/\~{}zgliczyn/papers/cc-3d/}.}\label{tab:sourceFiles}
\end{center}
\end{table}

%===================================
\section{Spatial central configurations} \label{sec:spatialCC}
The program finds all spatial configurations and writes them into output files. However, planar configurations have been already treated in~\cite{MZ}, thus here we present only non-coplanar CCs for $n=5$ and $n=6$.
We identify CCs presenting them with $U$, $J=U \sqrt{I}$ and the position of the before last body (recall that $q_{n-2} = (x_{n-2}, 0, 0)$)
 The geometric meaning of $x_{n-2}$ is: this is the distance of the body in CC which is furthers from the center of mass, hence this is another reasonable measure of the size of CC.
We present CCs also in figures so that their shape is more visible.

%------------------------------------------------
\subsection{Five bodies}
\label{sec:cc5b3d}
We prove that there exist  four classes  of  non-coplanar central configurations. This confirms  the results from \cite{LS,Kotsireas}. For each non-coplanar central configuration we find its all orthogonal symmetries.
These configuration are
\begin{itemize}
% CC6
\item diamond with triangular base Figure~\ref{fig:poly6}. Two symmetric pyramids with  an equilateral triangle  $q_0q_1q_2$  as the base, the summits of the pyramids $q_3$ and $q_4$ are on the axis perpendicular to the base plane passing through the center of mass. This solution in discussed in \cite[Sec. 5.2.3]{Kotsireas} as \emph{degree $12$ solution} and appears Fig. 3a in \cite{LS}
\begin{figure}[H]
\begin{minipage}[c]{0.6\textwidth}
\centering
\includegraphics[scale=0.5]{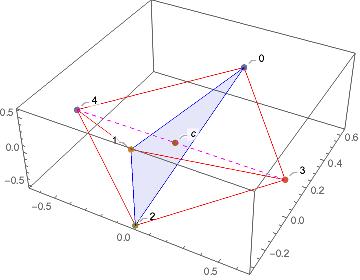}\\
\end{minipage}
\begin{minipage}[c]{0.2\textwidth}
\tiny
\begin{eqnarray*}
U & = & 0.4062917146317[56,92]\\[1ex]
 J & = & 0.2589744676461[55,92]\\[1ex]
 x_3 & = & 0.64538119213252[0,5]
 \end{eqnarray*}
 \end{minipage}
\caption{Diamond with triangular base, $n=5$}\label{fig:poly6}
\end{figure}

% CC7
\item a pyramid  with square base Figure~\ref{fig:poly7} , the square $q_0q_3q_2q_4$ with $q_1$ in the summit on the symmetry line; all triangular faces are equilateral;  $c$ indicates the center of mass and this lie over the plane of the square; diagonals of the square are marked as dashed blue lines.
This is called a \emph{square pyramid} in \cite[Sec. 5.2.1]{Kotsireas} and appears as Fig.  3b in \cite{LS}
\begin{figure}[H]
\begin{minipage}[c]{0.6\textwidth}
\centering
\includegraphics[scale=0.5]{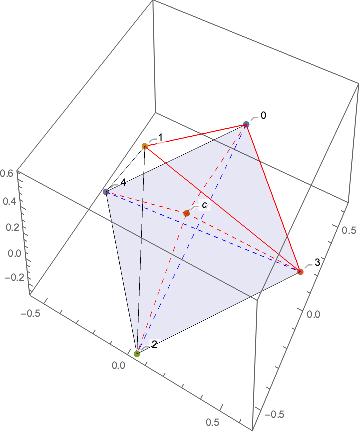}\\
\end{minipage}
\begin{minipage}[c]{0.2\textwidth}
\tiny
\begin{eqnarray*}
U & = & 0.4066684332090[57,96]\\[1ex]
 J & = & 0.2593347374996[26,70]\\[1ex]
 x_3 & = & 0.64031449289674[2,7]
 \end{eqnarray*}
 \end{minipage}
\caption{Pyramid  with square base, $n=5$}\label{fig:poly7}
\end{figure}

% CC8
\item the regular tetrahedron  (Figure~\ref{fig:poly8}) $q_0q_1q_3q_4$ with body $q_2$ at the origin, this is \emph{regular tetrahedron} from \cite[Sec. 5.2.2]{Kotsireas} and appears as   Fig. 4a in \cite{LS}
\begin{figure}[H]
\begin{minipage}[c]{0.6\textwidth}
\centering
\includegraphics[scale=0.5]{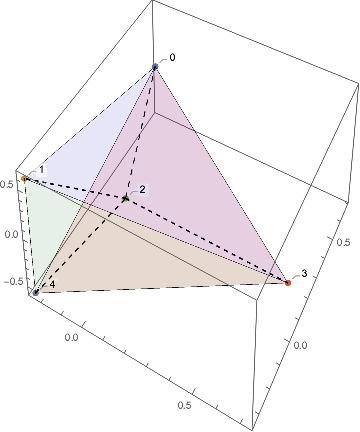}\\
\end{minipage}
\begin{minipage}[c]{0.2\textwidth}
\tiny
\begin{eqnarray*}
U & = & 0.4224351404453[00,38]\\[1ex]
 J & = & 0.2745617643612[04,48]\\[1ex]
 x_3 & = & 0.7266663096336[88,96]
 \end{eqnarray*}
 \end{minipage}
\caption{ The regular tetrahedron, $n=5$}\label{fig:poly8}
\end{figure}

% CC9
\item  triangle pyramid ("perturbed tetrahedron")  (Figure~\ref{fig:poly9}) with an equilateral triangle  $q_0q_1q_4$ as the base and $q_3$ at the summit; $q_2$ is inside;  c is at the origin. We identify it with the one  discussed in \cite[Sec. 5.2.4]{Kotsireas} as \emph{degree $43$ solution }, although
       no geometric description of the obtained central configuration has been given   there. It appears as Fig. 4b in \cite{LS}.
\begin{figure}[H]
\begin{minipage}[c]{0.6\textwidth}
\centering
\includegraphics[scale=0.5]{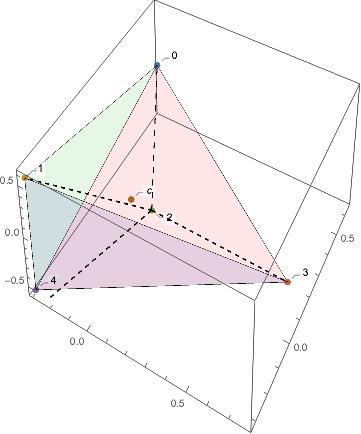}\\
\end{minipage}
\begin{minipage}[c]{0.2\textwidth}
\tiny
\begin{eqnarray*}
U & = & 0.4222317695727[56,83]\\[1ex]
 J & = & 0.2743635168460[37,68]\\[1ex]
 x_3 & = & 0.7779267783296[78,89]
 \end{eqnarray*}
 \end{minipage}
\caption{Perturbed tetrahedron, $n=5$}\label{fig:poly9}
\end{figure}

\end{itemize}

\subsection{Six bodies}\label{sec:six-presentation}
There are nine classes of non-coplanar central configurations:
\begin{enumerate}

% CC 10
\item diamond with a square base(Figure~\ref{fig:diamond}), two symmetrical pyramids  with a common square base $q_0q_4q_2q_5$; $q_1$ is the summit of the upper pyramid, $q_3$ --- the lower one. Points $q_1$, $q_3$ and $(0,0,0)$ (the center of mass) are collinear.
\begin{figure}[H]
\begin{minipage}[c]{0.6\textwidth}
\centering
\includegraphics[scale=0.55]{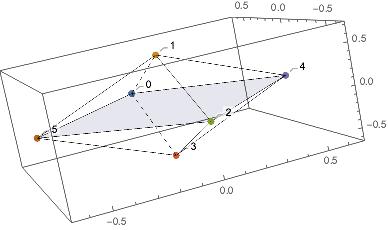}\\
\end{minipage}
\begin{minipage}[c]{0.2\textwidth}
\tiny
\begin{eqnarray*}
U & = & 0.4253096183510[01,17]\\[1ex]
 J & = & 0.2773689270621[74,91]\\[1ex]
x_4 & = & 0.6521576637217[46,51]
 \end{eqnarray*}
 \end{minipage}
\caption{Diamond with a square base (item 1), $n=6$}\label{fig:diamond}
\end{figure}

% CC 11
\item diamond with regular triangular base (Figure~\ref{fig:tetrahedron}), two tetrahedrons  with a common equilateral triangle base $q_0q_1q_3$; $q_2 = (0,0,0)$ lies at the  triangle plane (i.e.\ bodies $q_0$, $q_1$, $q_2$ and $q_3$ are coplanar); $q_4$ is the summit of right pyramid, $q_5$ --- the left one. Points $q_2$, $q_4$ and $q_5$ are collinear.
\begin{figure}[H]
\begin{minipage}[c]{0.6\textwidth}
\centering
\includegraphics[scale=0.5]{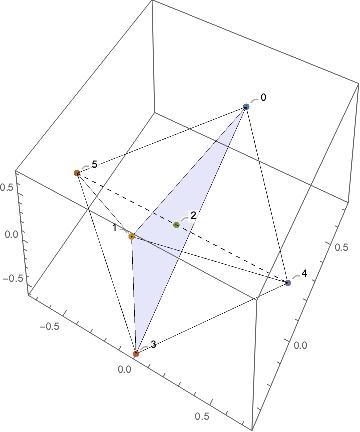}\\
\end{minipage}
\begin{minipage}[c]{0.2\textwidth}
\tiny
\begin{eqnarray*}
U & = & 0.439099368866[330,431]\\[1ex]
 J & = & 0.290967323607[179,293]\\[1ex]
x_4 & = & 0.728578780778[797,824]
 \end{eqnarray*}
 \end{minipage}
\caption{Diamond with regular triangular base (item 2), $n=6$}\label{fig:tetrahedron}
\end{figure}

% CC 12
\item two  pyramids (one inside the other) (Figure~\ref{fig:pyramid-12}) with a common square base $q_0q_1q_3q_5$; $q_2$ is the summit of inner pyramid, $q_4$ --- the outer one. Points $q_2$, $q_4$ and $(0,0,0)$ (the center of mass) are collinear.
\begin{figure}[H]
\begin{minipage}[c]{0.6\textwidth}
\centering
\includegraphics[scale=0.55]{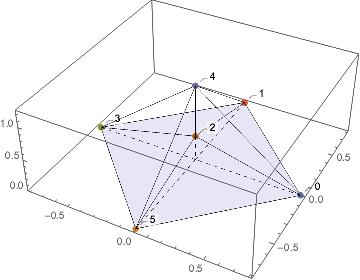}\\
\end{minipage}
\begin{minipage}[c]{0.2\textwidth}
\tiny
\begin{eqnarray*}
U & = & 0.43903620010[0984,1039]\\[1ex]
 J & = & 0.290904538096[071,137]\\[1ex]
x_4 & = & 0.7893559716313[63,73]
 \end{eqnarray*}
 \end{minipage}
\caption{ Two  pyramids (item 3), $n=6$ }\label{fig:pyramid-12}
\end{figure}

% CC 13
\item two orthogonal isosceles triangles (Figure~\ref{fig:airplane-13}); altitudes of both triangles, points $q_2$, $q_4$ and the center of mass (point $(0,0,0)$) lie on the line marked red on the picture, which is an intersection line of the planes containing the triangles
\begin{figure}[H]
\begin{minipage}[c]{0.6\textwidth}
\centering
\includegraphics[scale=0.5]{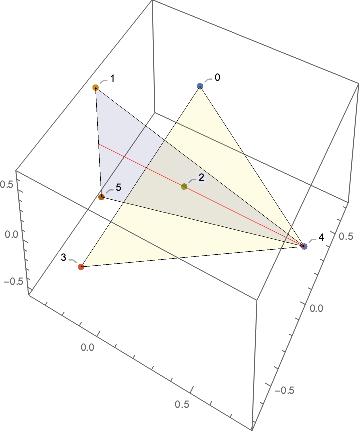}\\
\end{minipage}
\begin{minipage}[c]{0.2\textwidth}
\tiny
\begin{eqnarray*}
U & = & 0.4388961044945[10,57]\\[1ex]
 J & = & 0.290765308587[179,232]\\[1ex]
x_4 & = & 0.7815214369079[53,63]
 \end{eqnarray*}
 \end{minipage}
\caption{ Two orthogonal isosceles triangles (item 4) $n=6$}\label{fig:airplane-13}
\end{figure}

% CC 14
\item two  pyramids (one inside the other) (Figure~\ref{fig:pyramid-14}) with a common square base $q_0q_1q_3q_5$; $q_2$ is the summit of inner pyramid, $q_4$ --- the outer one. Points $q_2$, $q_4$ and $(0,0,0)$ (the center of mass) are collinear.
\begin{figure}[H]
\begin{minipage}[c]{0.6\textwidth}
\centering
\includegraphics[scale=0.5]{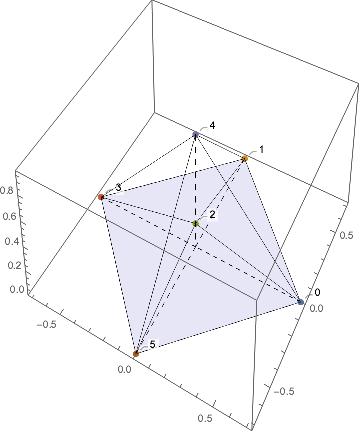}\\
\end{minipage}
\begin{minipage}[c]{0.2\textwidth}
\tiny
\begin{eqnarray*}
U & = & 0.439337329879[509,619]\\[1ex]
 J & = & 0.291203881424[758,879]\\[1ex]
x_4 & = & 0.7268223700720[89,95]
 \end{eqnarray*}
 \end{minipage}
\caption{Two pyramids (item 5), $n=6$}\label{fig:pyramid-14}
\end{figure}

% CC 15
\item pentagonal pyramid, (Figure~\ref{fig:pyramid-15}) a polyhedron with a  regular pentagon $q_0q_4q_2q_5q_3$ as a base; $q_1$ is the summit and is collinear with $(0,0,0)$ (the center of mass).
\begin{figure}[H]
\begin{minipage}[c]{0.6\textwidth}
\centering
\includegraphics[scale=0.5]{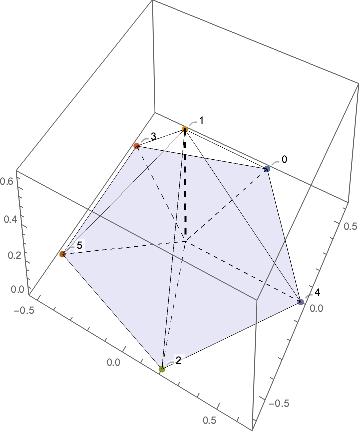}\\
\end{minipage}
\begin{minipage}[c]{0.2\textwidth}
\tiny
\begin{eqnarray*}
U & = & 0.4327554322378[11,42]\\[1ex]
 J & = & 0.2846844804120[32,68]\\[1ex]
x_4 & = & 0.66271482439001[3,9]
 \end{eqnarray*}
 \end{minipage}
\caption{Pentagonal pyramid (item 6), $n=6$ }\label{fig:pyramid-15}
\end{figure}

% CC 16
\item a prism (Figure~\ref{fig:pyramid-16}), a polyhedron with three rectangular  and two triangular faces; $q_0q_1q_4$ and $q_2q_3q_5$ are symmetrical equilateral triangles, thus rectangles $q_0q_1q_3q_5$, $q_0q_4q_2q_5$ and $q_1q_3q_2q_4$ are also of the same size --- lines with the same length have the same color (red or blue).
\begin{figure}[H]
\begin{minipage}[c]{0.6\textwidth}
\centering
\includegraphics[scale=0.6]{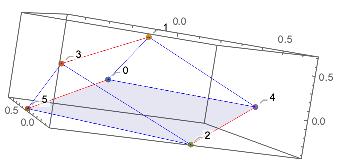}\\
\end{minipage}
\begin{minipage}[c]{0.2\textwidth}
\tiny
\begin{eqnarray*}
U & = & 0.428444021033[087,122]\\[1ex]
 J & = & 0.280440756454[764,805]\\[1ex]
x_4 & = & 0.6545563543600[36,42]
 \end{eqnarray*}
 \end{minipage}
\caption{Prism (item 7), $n=6$}\label{fig:pyramid-16}
\end{figure}

% CC 17
\item triangular polyhedron (Figure~\ref{fig:pyramid-17}) with  faces being isosceles triangles; $q_0q_1q_4$ and $q_0q_1q_5$ are symmetrical, $q_0q_4q_5$ has one side longer (the segment $(q_4, q_5)$ is longer than $(q_0,q_1)$); triangle $q_0q_2q_3$ is also an isosceles triangle;  lines with the same length have the same color (red or blue).
\begin{figure}[H]
\begin{minipage}[c]{0.6\textwidth}
\centering
\includegraphics[scale=0.55]{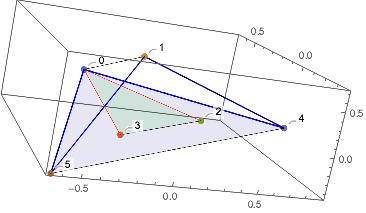}\\
\end{minipage}
\begin{minipage}[c]{0.2\textwidth}
\tiny
\begin{eqnarray*}
U & = & 0.4583877507512[24,44]\\[1ex]
 J & = & 0.3103483962997[44,68]\\[1ex]
x_4 & = & 0.88496678932040[3,7]
 \end{eqnarray*}
 \end{minipage}
\caption{Triangular polyhedron (item 8), $n=6$}\label{fig:pyramid-17}
\end{figure}

% CC 18
\item Diamond with triangular base, (Figure \ref{fig:pyramid-18}) two symmetrical polyhedrons with triangular base $q_1q_4q_5$ and with summits at $q_0$ and $q_3$; the center of mass ($(0,0,0)$) and the point $q_2$ lies on the plane of triangle $q_1q_4q_5$ and points $q_0$, $q_3$ are symmetrical with respect to that plane (i.e.\ triangle $q_0q_2q_3$ is isosceles); in the figure point $c$ is the center of mass $(0,0,0)$;  lines with the same length have the same color (red, blue, magenta, green and orange).
\begin{figure}[H]
\begin{minipage}[c]{0.6\textwidth}
\centering
\includegraphics[scale=0.5]{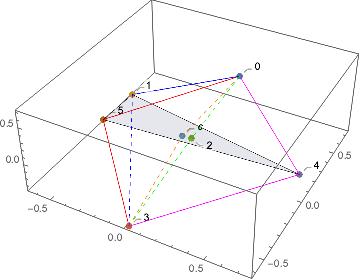}\\
\end{minipage}
\begin{minipage}[c]{0.2\textwidth}
\tiny
\begin{eqnarray*}
U & = & 0.439320318180[620,757]\\[1ex]
 J & = & 0.291186967912[789,943]\\[1ex]
x_4 & = & 0.7520348731739[36,75]
 \end{eqnarray*}
 \end{minipage}
\caption{Diamond with isosceles triangular base (item 9) $n=6$}\label{fig:pyramid-18}
\end{figure}

\end{enumerate}

%===================================
\section{The number of non-equivalent  CCs for mass parameters close to equal mass case}
\label{sec:iso}

In this section we count the number of CCs (different equivalency classes of CCs) for mass parameters close to the equal mass case.
This is possible because for the equal mass case all CCs turned out to be non-degenerate solutions of \RS\!\!.
 Then a simple
continuation argument allows us to infer that the number of CCs does not change.
A  single CC for the equal mass case give rise (can be continued) to multiple
CCs when the masses differ.

\begin{example}\rm
Consider a square --- one of the planar CCs for $n=4$.
We distinguish bodies by (possibly different) masses  and we show how many CCs
for different masses we obtain. We use vertex labeling (coloring) for this purpose.
To suggest different arrangement of masses we use $m_1,\ldots, m_4$ colors. We identify those colorings which result from the application
of planar isometry to  a CC.
In Figure~\ref{fig:square} we present all three different  arrangements of bodies for CC being a square under symmetry group $\mathcal{O}(2)$.
 Notice that there are 4! = 24 different colorings of a square by four colors $m_1, \ldots, m_4$. However, some of them can be obtained from another by rotation or reflection. For example arrangement $m_4m_1m_2m_3$ can be obtained from $m_1m_2m_3m_4$ by rotating latter by $\pi/2$ angle.
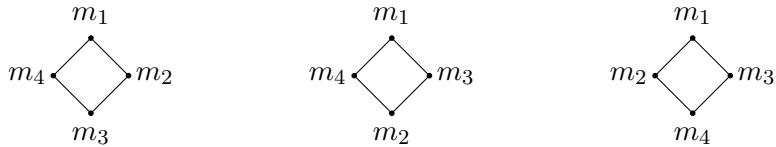
\begin{figure}[htb]
  \centering
\begin{tikzpicture}
% (1234)
\draw[fill] (0,0) circle (0.03cm);
\node[] at (0.0, 0.3)  {$m_1$};

\draw[fill] (0.5,-0.5) circle (0.03cm);
\node[] at (0.85, -0.5)  {$m_2$};

\draw[fill] (-0.5,-0.5) circle (0.03cm);
\node[] at (-0.85, -0.5)  {$m_4$};
\draw[] (-0.5, -0.5) -- (0.0, 0.0);
\draw[] (0.0, 0.0) -- (0.5, -0.5);

\draw[fill] (0,-1.0) circle (0.03cm);
\node[] at (0.0, -1.3)  {$m_3$};
\draw[] (0.5, -0.5) -- (0.0, -1.0);
\draw[] (0.0, -1.0) -- (-0.5, -0.5);

% (1342)
\draw[fill] (4,0) circle (0.03cm);
\node[] at (4.0, 0.3)  {$m_1$};

\draw[fill] (4.5,-0.5) circle (0.03cm);
\node[] at (4.85, -0.5)  {$m_3$};

\draw[fill] (3.5,-0.5) circle (0.03cm);
\node[] at (3.15, -0.5)  {$m_4$};
\draw[] (3.5, -0.5) -- (4.0, 0.0);
\draw[] (4.0, 0.0) -- (4.5, -0.5);

\draw[fill] (4,-1.0) circle (0.03cm);
\node[] at (4.0, -1.3)  {$m_2$};
\draw[] (4.0, -1.0) -- (3.5, -0.5);
\draw[] (4.0, -1.0) -- (4.5, -0.5);

% (1423)
\draw[fill] (8,0) circle (0.03cm);
\node[] at (8.0, 0.3)  {$m_1$};

\draw[fill] (8.5,-0.5) circle (0.03cm);
\node[] at (8.85, -0.5)  {$m_3$};

\draw[fill] (7.5,-0.5) circle (0.03cm);
\node[] at (7.15, -0.5)  {$m_2$};
\draw[] (7.5, -0.5) -- (8.0, 0.0);
\draw[] (8.0, 0.0) -- (8.5, -0.5);

\draw[fill] (8,-1.0) circle (0.03cm);
\node[] at (8.0, -1.3)  {$m_4$};
\draw[] (8.0, -1.0) -- (7.5, -0.5);
\draw[] (8.0, -1.0) -- (8.5, -0.5);
\end{tikzpicture}
\caption{All non-isomorphic  (non-$\mathcal{O} (2)$ equivalent) colorings of a square with colors $m_1, \ldots, m_4$.}\label{fig:square}
\end{figure}

If we  use $\mathcal{SO}(2)$ as the symmetry group (we allow  rotations and reject reflections) we obtain six colorings, i.e.\ six non-congruous CCs (see Figure~\ref{fig:square-con}).
\begin{figure}[htb]
  \centering
\begin{tikzpicture}
% (1234)
\draw[fill] (0,3) circle (0.03cm);
\node[] at (0.0, 3.3)  {$m_1$};

\draw[fill] (0.5,2.5) circle (0.03cm);
\node[] at (0.85, 2.5)  {$m_2$};

\draw[fill] (-0.5,2.5) circle (0.03cm);
\node[] at (-0.85, 2.5)  {$m_4$};
\draw[] (-0.5, 2.5) -- (0.0, 3.0);
\draw[] (0.5, 2.5) -- (0.0, 3.0);

\draw[fill] (0,2.0) circle (0.03cm);
\node[] at (0.0, 1.7)  {$m_3$};
\draw[] (0.5, 2.5) -- (0.0, 2.0);
\draw[] (0.0, 2.0) -- (-0.5, 2.5);

% (1342)
\draw[fill] (4,3) circle (0.03cm);
\node[] at (4.0, 3.3)  {$m_1$};

\draw[fill] (4.5,2.5) circle (0.03cm);
\node[] at (4.85, 2.5)  {$m_3$};

\draw[fill] (3.5,2.5) circle (0.03cm);
\node[] at (3.15, 2.5)  {$m_4$};
\draw[] (3.5, 2.5) -- (4.0, 3.0);
\draw[] (4.0, 3.0) -- (4.5, 2.5);

\draw[fill] (4,2.0) circle (0.03cm);
\node[] at (4.0, 1.7)  {$m_2$};
\draw[] (4.0, 2.0) -- (3.5, 2.5);
\draw[] (4.0, 2.0) -- (4.5, 2.5);

% (1423)
\draw[fill] (8,3) circle (0.03cm);
\node[] at (8.0, 3.3)  {$m_1$};

\draw[fill] (8.5,2.5) circle (0.03cm);
\node[] at (8.85, 2.5)  {$m_3$};

\draw[fill] (7.5,2.5) circle (0.03cm);
\node[] at (7.15, 2.5)  {$m_2$};
\draw[] (7.5, 2.5) -- (8.0, 3.0);
\draw[] (8.0, 3.0) -- (8.5, 2.5);

\draw[fill] (8,2.0) circle (0.03cm);
\node[] at (8.0, 1.7)  {$m_4$};
\draw[] (8.0, 2.0) -- (7.5, 2.5);
\draw[] (8.0, 2.0) -- (8.5, 2.5);

%
%1432
\draw[fill] (0,0) circle (0.03cm);
\node[] at (0.0, 0.3)  {$m_1$};

\draw[fill] (0.5,-0.5) circle (0.03cm);
\node[] at (0.85, -0.5)  {$m_4$};

\draw[fill] (-0.5,-0.5) circle (0.03cm);
\node[] at (-0.85, -0.5)  {$m_2$};
\draw[] (-0.5, -0.5) -- (0.0, 0.0);
\draw[] (0.0, 0.0) -- (0.5, -0.5);

\draw[fill] (0,-1.0) circle (0.03cm);
\node[] at (0.0, -1.3)  {$m_3$};
\draw[] (0.5, -0.5) -- (0.0, -1.0);
\draw[] (0.0, -1.0) -- (-0.5, -0.5);

% (1423)
\draw[fill] (4,0) circle (0.03cm);
\node[] at (4.0, 0.3)  {$m_1$};

\draw[fill] (4.5,-0.5) circle (0.03cm);
\node[] at (4.85, -0.5)  {$m_4$};

\draw[fill] (3.5,-0.5) circle (0.03cm);
\node[] at (3.15, -0.5)  {$m_3$};
\draw[] (3.5, -0.5) -- (4.0, 0.0);
\draw[] (4.0, 0.0) -- (4.5, -0.5);

\draw[fill] (4,-1.0) circle (0.03cm);
\node[] at (4.0, -1.3)  {$m_2$};
\draw[] (4.0, -1.0) -- (3.5, -0.5);
\draw[] (4.0, -1.0) -- (4.5, -0.5);

% (1243)
\draw[fill] (8,0) circle (0.03cm);
\node[] at (8.0, 0.3)  {$m_1$};

\draw[fill] (8.5,-0.5) circle (0.03cm);
\node[] at (8.85, -0.5)  {$m_2$};

\draw[fill] (7.5,-0.5) circle (0.03cm);
\node[] at (7.15, -0.5)  {$m_3$};
\draw[] (7.5, -0.5) -- (8.0, 0.0);
\draw[] (8.0, 0.0) -- (8.5, -0.5);

\draw[fill] (8,-1.0) circle (0.03cm);
\node[] at (8.0, -1.3)  {$m_4$};
\draw[] (8.0, -1.0) -- (7.5, -0.5);
\draw[] (8.0, -1.0) -- (8.5, -0.5);

\end{tikzpicture}
\caption{All non-congruous (not $\mathcal{SO}(2)$-equivalent)  colorings of a square with colors $m_1, \ldots, m_4$.}\label{fig:square-con}
\end{figure}
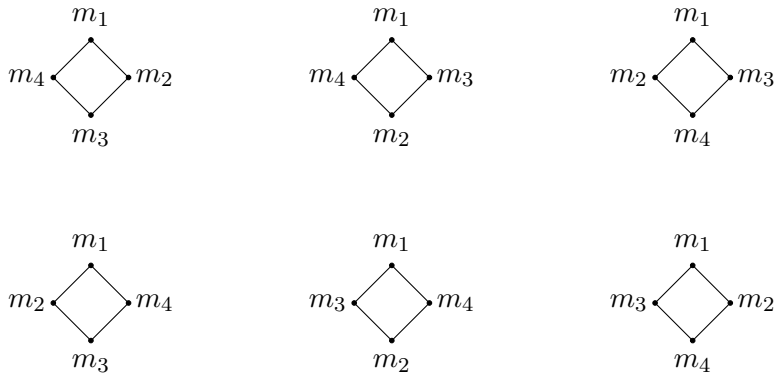

 Summarizing: if we use just rotations, then there is six non-equivalent CCs obtained from the square  in the different masses case; if we additionally allow reflections, then we obtain only three non-equivalent CCs.
\end{example}

%------------------------------------------------
\subsection{Definitions and P{\'o}lya's theorem}

We can count non-isomorphic colorings with the aid of P{\'o}lya's enumeration theorem, see~\cite{B89} or~\cite{PTW83}) for the detailed treatment and proof. In this section we just recall the relevant definitions and state main theorems.

Let $X$ be a finite set with $|X| = n$ and $(G, \circ)$ be a subgroup of the group of  permutations of $X$ with $\circ$ denoting the composition of permutations.

Coloring of $X$ is  a function $\omega: X\to \mathcal{C}$, where $ \mathcal{C}$  is a finite set of colors. We assume that the cardinality of $\mathcal{C}$ is $k$ and it is the only important feature of $\mathcal{C}$.
Notice that  any non-negative $k$ has a combinatorial sense, but in our counting problem we have to take $k = n$.

\begin{definition}
\label{def:col-iso}
We say that two colorings $\omega_1$, $\omega_2$ are {\em isomorphic}  with respect to group $G$, if there is a $g\in G$ such that
$$\omega_1(g(x)) = \omega_2(x), \quad \forall x\in X.$$
\end{definition}

\begin{definition}
An {\em index of a permutation} $g\colon X\to X$ is a function of $n$ variables
$$\zeta_{g}\left( x_1,\ldots,x_n \right) =x_1^{\alpha_1}\cdot x_2^{\alpha_2}\cdot\ldots\cdot x_n^{\alpha_n},$$
where
$\alpha_i$ is a number of  cycles  of permutation $g$ containing exactly $i$ elements
for $i=1,2,\ldots, n.$
\end{definition}

\begin{definition}
An {\em index of a group} $G$ is defined as
$$
\zeta_G(x_1, \ldots, x_n) =
 \frac{1}{|G|} \sum_{g\in G}\zeta_{g}(x_1, \ldots, x_n).
$$
\end{definition}

\begin{theorem}
The number of non-isomorphic colorings of $X$  with respect to group $G$ with $k$ colors is
$$\zeta_{G}\left( k,k,\ldots,k \right).$$
\end{theorem}

\begin{example}\label{ex:square2}\rm
Let $X = \{1 , 2,  3,  4\}$ be a square as in Figure~\ref{fig:square}.  Let $(G_{\mbox{\sc\tiny RR}}, \circ)$ be a group of all rotations and
reflections (on a plane) of that square.
Permutations in $G_{\mbox{\sc\tiny RR}}$, their decomposition in two cycles and their indices are:\\[1ex]
$
\begin{array}{l|l|lcl|p{75mm}}
g_0 &  (1)(2)(3)(4) & \zeta_{g_0}(x_1, x_2, x_3, x_4) & = & x_1^4 &  identity \\
g_1 & (1234) & \zeta_{g_1}(x_1, x_2, x_3, x_4) & = & x_4 &   rotation by $\pi/2$ \quad (one 4-elements cycle)\\
g_2 & (13)(24) & \zeta_{g_2}(x_1, x_2, x_3, x_4) & = & x_2^2 & rotation by $\pi$ \quad (two 2-elements cycles)\\
g_3 & (1432) &\zeta_{g_3}(x_1, x_2, x_3, x_4) & = & x_4 &    rotation by $3\pi/2$ \quad (one 4-elements cycle)\\
g_4 & (1)(3)(24) & \zeta_{g_4}(x_1, x_2, x_3, x_4) & = & x_1^2  x_2 &  reflection with respect to line (13) \quad (two 1-element cycles, one 2-element cycle)\\
g_5 & (13)(2)(4) & \zeta_{g_5}(x_1, x_2, x_3, x_4) & = & x_1^2  x_2 &  reflection with respect to line (24)\\
g_6 & (12)(34) & \zeta_{g_6}(x_1, x_2, x_3, x_4) & = & x_2^2 &  reflection with respect to line perpendicular to the segment (12)\quad (two 2-elements cycles)\\
g_7 & (14)(23) & \zeta_{g_7}(x_1, x_2, x_3, x_4) & = & x_2^2 & reflection with respect to line perpendicular to the segment (14).
\end{array}
$
\mbox{}\\[1ex]
Thus $G_{\mbox{\sc\tiny RR}} = \{g_0, \ldots, g_7\}$, $|G_{\mbox{\sc\tiny RR}}| = 8$ and its index is
$$\zeta_{G_{\mbox{\sc\tiny RR}} } (x_1, x_2, x_3, x_4) = \frac{1}{8}\left(x_1^4 + 3x_2^2 + 2x_4 + 2x_1^2x_2\right).
$$
Hence the number of different coloring with $4$ colors is
$$\zeta_{G_{\mbox{\sc\tiny RR}} } (4,4,4,4) = \frac{1}{8}\left(4^4 + 4^2 + 2\cdot 4\right) = 2360.
$$
In this number there are all colorings with one, two, three and four colors, and this is not   what we are interested in. We want to know a number of different colorings with exactly four colors. For this we need full version of P{\'o}lya's enumeration Theorem.
\end{example}

\begin{theorem}[P{\'o}lya's enumeration Theorem]
Let $D$ be a set of all non-isomorphic colorings of a set $X$  with respect to $G$  with $k$ colors. Then the generating function $\mathcal{U}_{D}$
$$\mathcal{U}_{D}\left( x_1,x_2,\ldots,x_k \right)=\zeta_{G}\left( \sigma_1,\sigma_2,\ldots,\sigma_n \right),$$
where
$$\sigma_i=x_1^i+x_2^i+\ldots+x_k^i,$$
has its coefficient at $x_1^{i_1}\ldots x_k^{i_k}$ equal to the number of non-isomorphic colorings using each of the colors $m_1, \ldots, m_k$ exactly $i_1, \ldots, i_k$ times, respectively.
\end{theorem}

\begin{example}\rm
Consider the case from Example~\ref{ex:square2}. The generating function is
\begin{eqnarray}
\mathcal{U}(x_1, x_2, x_3, x_4) & = & \zeta_{G_{\mbox{\sc\tiny RR}}}(x_1+x_2+x_3+x_4, x_1^2+x_2^2+x_3^2+x_4^2, \nonumber\\
 & & \quad\quad\quad x_1^3+x_2^3+x_3^3+x_4^3, x_1^4+x_2^4+x_3^4+x_4^4)
\nonumber\\
 & = &\frac{1}{8}
   \Big\{\ (x_1+x_2+x_3+x_4)^4 +
   \nonumber\\
  &  &  \quad\quad 3(x_1^2 + x_2^2 + x_3^2 + x_4^2)^2 +
   \nonumber\\
 &  & \quad\quad  2(x_1^4 + x_2^4 + x_3^4 +2x_4^4) +
  \nonumber\\
 &  & \quad\quad 2(x_1 + x_2 + x_3 + x_4)^2(x_1^2 + x_2^2 + x_3^2 + x_4^2)\Big\}\nonumber\\
 & = & x_1^4 + \ldots + \textbf{3}x_1x_2x_3x_4 + \ldots x_4^4\label{eqn:no-colors}
 \end{eqnarray}
 From the expansion above~(\ref{eqn:no-colors}), we see that the number of non-isomorphic colorings with four colors, each used once, in the case of square is 3.
\end{example}

%------------------------------------------------
\subsection{Identifications of CCs made for the main result --- summary}
Let us remind the essence of counting of CCs mentioned in  Section~\ref{sec:main-result}:  any CC can be obtained from one of the normalized CCs, found by the program, by a suitable composition of translation, scaling, rotation, reflection and permutation of bodies. The reasons for such situation are as follows.
\begin{enumerate}
\item First, in Section~\ref{sec:cc-eq}, we normalized central configurations to obtain isolated solutions. This means that we remove possibility of translation (establishing a center of mass $c=0$) and possibility of scaling symmetry (setting $\lambda = 1$). However, the obtained system~(\ref{eq:cc-kart}) can have $\mathcal{O}(3)$ and $\mathcal{SO}(3)$ symmetry, which excludes the use of Krawczyk's method. This method requires non-degenerate solutions .

\item In Section~\ref{sec:red-sys}, we introduce \RS and after these treatments solutions found by our program have no longer neither $\mathcal{SO}(3)$  nor $\mathcal{O}(3)$ symmetry. But reflections by the planes and some permutations of bodies are still possible. In the case of equal masses two CCs that differ only by labeling of bodies are equivalent.

\item Thus we  remove remaining symmetries and possible permutations of bodies by a procedure of unifications of solutions.
\end{enumerate}
Thus, in the equal mass case we count configurations treating them as indistinct if they have the same geometrical form. In the context of the P{\'o}yla's theorem this corresponds to taking the permutation group $G$ to be the whole of $\mathcal{S}_n$ or, equivalently, using just one color.

%------------------------------------------------
\subsection{Number of CCs for different masses close to equal mass case}

In the different masses case in the context of the P{\'o}yla's theorem we use exactly $n$ colors.
We consider two CCs equivalent,  \emph{isomorphic} or  \emph{congruous}, if one can be transformed into another by an element of $\mathcal{O}(d)$ or $\mathcal{SO}(d)$, respectively.
Hence depending on whether we allow for reflections (i.e.\ $\mathcal{O}(d)$) or not ($\mathcal{SO}(d)$) we get different counts.

In the sequel by  no-CC$(n)$ we denote the number of CCs obtained for the equal masses case, by $\mbox{iso}(n)$ is number of different  non-isomorphic CCs and  the number of non-congruous CCs is denoted by $\homo(n)$. These numbers are obtained  using P{\'o}lya's enumeration Theorem for different groups.
Tables~\ref{tab:homo2D} and~\ref{tab:homo3D} contain the  number of different CCs inferred from our rigorous count of CCs. Data in Tables~\ref{tab:homo2D} and~\ref{tab:homo3D}  suggest that the number of different CCs in the different mass case grow
faster than $(n!)$.

\begin{example}\rm
For  $n= 4$ in the planar case there are four different CCs (i.e.\ no-CC(4) = 4) and there are
$12$ non-isomorphic configurations for collinear solution, $3$  for square,
 $4$ for equilateral triangle and  $12$ for isosceles triangle
(for detailed description of CCs see~\cite{MZ}), thus iso$(4) = 12 + 3 + 12 + 4 = 31$. The count of non-congruous classes is as follows:
$12$ non-congruous configurations for collinear solution, $6$  for square,
 $8$ for equilateral triangle and  $24$ for isosceles triangle, thus $\homo(4)=12+6+8+24=50$.
\end{example}

\begin{table}[H]
\begin{center}
\begin{tabular}{c|c|c|c|c|c}
$n$ & no-CC$(n)$ & iso$(n)$ & $\homo(n)$ & iso$(n)/n!$ & $\homo(n)/n!$\\
\hline
4 & 4 & 31 & 50 & 1.29167 & 2.0833\\
5 & 5 & 207 & 354 &  1.72500 & 2.9500\\
6 & 9 & 1992 & 3624 & 2.76667 & 5.0333\\
7 & 14 & 28080 & 53640 & 5.57143 & 10.6429
\end{tabular}
\end{center}
\caption{The number of different CCs  in 2D}\label{tab:homo2D}
\end{table}

\begin{table}[H]
\begin{center}
\begin{tabular}{c|c|c|c|c|c|}
$n$ & no-CC$(n)$ & iso$(n)$ & $\homo(n)$  &  iso$(n)/n!$ & $\homo(n)/n!$\\
\hline
4 & 5 & 32 & 52 & 1.33333 & 2.1667\\
5 & 9 & 257 & 454 & 2.14167 & 3.7833\\
6 & 18 & 3099 & 5838 & 4.30417 & 8.1083
\end{tabular}
\end{center}
\caption{The number of different CCs in 3D}\label{tab:homo3D}
\end{table}

%===================================
\section{The question of stability for planar CCs}
\label{sec:stable-planar}

Now, consider a planar CC.  It gives rise to a periodic orbit, where all bodies move on circles with the angular velocity $1$.
This orbit becomes a fixed point in the rotating coordinate frame.  The stability/instability of this fixed point and the circular periodic orbit
is the same.

Let
\begin{equation}
  J=\left[
    \begin{array}{cc}
      0 & -1 \\
      1 & 0 \\
    \end{array}
  \right].
\end{equation}
Then $\exp(J\theta)=\left[
    \begin{array}{cc}
      \cos \theta & -\sin \theta \\
      \sin \theta & \cos \theta \\
    \end{array}
  \right]$
is a rotation by the angle $\theta$ in the plane $OXY$.

The link between coordinates in the inertial frame $x \in \mathbb{R}^2 $ and the coordinates $q$ with respect to the rotating frame with
the angular velocity equal to $1$ is
\begin{equation}
  x=\exp(Jt)q.
\end{equation}

The equations of motion in the rotating coordinate frame are  \cite{Si78}
\begin{eqnarray}
  \dot{q}_i&=&v_i,  \label{eq:qi=vi} \\
  \dot{v}_i&=& -2 Jv_i + q_i - \sum_{j \neq i} \frac{m_j (q_i - q_j)}{r_{ij}^3}. \label{eq:v-i'}
\end{eqnarray}

As was mentioned earlier each planar CC with $v_i=0$, $i=0,\dots,n-1$ is a fixed point of the system (\ref{eq:qi=vi},\ref{eq:v-i'}).  We investigated numerically the linear stability of all CCs for $n=4,5,6,7$ and for
some particular non-symmetric  CCs for $n=8,9,10$ whose existence has been established in
 \cite{MZ}. It  turns out that all these CCs are linearly unstable. The computation of eigenvalues
 for the linearization of  (\ref{eq:qi=vi},\ref{eq:v-i'}) has been done non-rigorously, but we are confident that
 this computation can be with some effort made rigorous.

%===================================

\end{document}